\documentclass[prl, twocolumn, showpacs,preprintnumbers,amsmath,amssymb,tightenlines,epsfig,floatfix]{revtex4}

\usepackage{graphicx}
\usepackage{dcolumn}
\usepackage{bm}
\usepackage{amsfonts}

\begin{document}
\preprint{Version: submitted \today}
\title{Generating controllable atom-light entanglement with a Raman atom laser system }
\author{S.A. Haine$^1$, M. K. Olsen$^2$ and J.J. Hope$^1$}
\affiliation{$^1$ Australian Centre for Quantum-Atom Optics, The Australian National University, Canberra, 0200, Australia.
$^2$ Australian Centre for Quantum-Atom Optics, University of Queensland, Brisbane, 4072, Australia.}

\begin{abstract}
We introduce a scheme for creating continuous variable entanglement between an atomic beam and an optical field, by using squeezed light to outcouple atoms from a BEC via a Raman transition. We model the full multimode dynamics of the atom laser beam and the squeezed optical field, and show that with appropriate two-photon detuning and two-photon Rabi frequency, the transmitted light is entangled in amplitude and phase with the outcoupled atom laser beam. The degree of entanglement is controllable via changes in the two-photon Rabi frequency of the outcoupling process.
\end{abstract}


\maketitle
{\it Introduction:}
Interferometry using massive particles promises hugely increased sensitivity over that available optically \cite{Rolston}. As one example, given equal enclosed area and particle flux, the sensitivity of atom interferometer gyroscopes exceeds that of photonic gyroscopes by a factor of $10^{11}$  \cite{Gustavson}. A seemingly obvious route to take advantage of this feature is the use of atom lasers in interferometry. However, there is a problem in that their flux cannot be increased arbitrarily due to the non-Markovian nature of the outcoupling process \cite{robins}. Hence there is much interest in finding alternate methods for improving the sensitivity.  As optical detection technology is more developed than that used for atomic detection, it seems  that it would be advantageous to adapt atom-light interferometers such as that demonstrated recently at MIT \cite{ketterle_interferometer} to combine the sensitivity of atoms to rotations, gravitational and magnetic fields, with the convenience and detection efficiency of light. Having previously shown that a quadrature squeezed atom laser can be made by using squeezed light to outcouple atoms from a trapped Bose-Einstein condensate (BEC) \cite{squeezy1,squeezy2}, we now introduce a scheme which produces entanglement between the outcoupled atoms and the transmitted light. This scheme could potentially be used to increase the sensitivity of an atom-light interferometer to below the standard quantum limit, and approach the Heisenberg limit \cite{Dowling}. 

Although schemes using BEC to entangle atoms and light have previously been proposed \cite{PierreMeystre}, both the atom laser output and the optical output of our scheme are directional, controlled and can have a reasonable flux. The continuous variable entanglement present also facilitates the inference of quantum statistical properties of the atomic beam via homodyne measurements on the output optical field, without needing the complication of atomic homodyne measurements. 

\begin{figure}
\includegraphics[width=5cm]{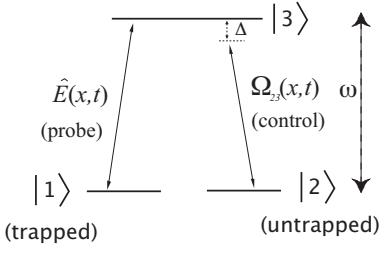}
\caption{\label{fig:levels} Internal energy levels of our three-level atom. A condensate of state $|1\rangle$ atoms confined in a trapping potential is coupled to free space via a Raman transition.  The two fields of the Raman transition are a probe beam (annihilation operator $\hat{E}(x,t)$) and a control field, which we have assumed is strong compared to the probe beam and can be approximated by a classical field ($\Omega_{23}(x,t)$) which is detuned from the excited state by an amount $\Delta$.}
\end{figure}

Our scheme is based on a Raman atom laser (Fig \ref{fig:levels}). A BEC consisting of three-level atoms is confined in a magnetic trap and manipulated by two laser fields. A photon from the probe beam is absorbed, and one is emitted into the control beam, transferring the internal state of the atom from $|1\rangle$ (trapped) to $|2\rangle$ (untrapped) and giving the atom a momentum kick of $\hbar(\mathbf{k}_{\mbox{probe}} - \mathbf{k}_{\mbox{control}})$, forming an atom laser beam. We showed in \cite{squeezy1, squeezy2} that, under appropriate conditions, the quantum state of the probe field can be transferred almost completely to the atom laser beam. The feature we stress in this letter is that when the two-photon Rabi frequency is less than optimal for complete quantum state transfer, the initial quantum state of the probe field is shared with the atomic beam, resulting in continuous variable entanglement between the two. 
This is reminiscent of a variable reflectivity beam splitter with an arbitrary quantum state at one input, and vacuum at the other. It has previously been shown that a 50/50 beam splitter with quadrature squeezed light entering one port and vacuum at the other port yields continuous variable entanglement in amplitude and phase at the two outputs \cite{bias-entangle}. The model we develop here is a little more complicated because we must consider the full multimode dynamics of the optical and atomic fields to take into account the absorption of the probe beam as it travels through the condensate and the possibility of outcoupled atoms coupling back in to the condensate due to the finite time they spend in the interaction region, but essentially operates on the same principle. 

The Hamiltonian describing the system is
\begin{eqnarray}
\mathcal{H} &=& \mathcal{H}_{\mbox{atom}} + \mathcal{H}_{\mbox{int}}  + \mathcal{H}_{\mbox{light}} \\ \nonumber
&=& \int \hat{\psi}^{\dag}_1(x) H_1 \hat{\psi}_1(x) dx + \int \hat{\psi}^{\dag}_2(x)(-\frac{\hbar^2}{2m}\nabla^2)  \hat{\psi}_2(x) dx \\ \nonumber
 &+& \int \hat{\psi}^{\dag}_3(x)(-\frac{\hbar^2}{2m}\nabla^2 + \hbar\omega_0) \hat{\psi}_3(x) dx \\ \nonumber
&+& \hbar\int(\hat{\psi}_2(x)\hat{\psi}^{\dag}_3(x)\Omega(x,t) + h.c.)dx \\ \nonumber
&+& \hbar g_{13}\int(\hat{E}(x)\hat{\psi}_1(x)\hat{\psi}^{\dag}_3 + h.c.) dx  + \mathcal{H}_{\mbox{light}} ,
\end{eqnarray}
where $H_1 = -\frac{\hbar^2}{2m}\nabla^2 + V_{trap}(x)$ , $\Omega_{23}(x,t) = \Omega_{23} e^{i(k_0 x - (\omega_0 - \Delta)t)}$ where $\Omega_{23}$ is the Rabi frequency for the $|2\rangle \rightarrow |3\rangle$ transition, and $m$ is the mass of the atoms.  $\hat{\psi}_{1}(x)$, $\hat{\psi}_{2}(x)$, $\hat{\psi}_{3}(x)$ and $\hat{E}(x)$ are the annihilation operators for the condensate mode (internal state $|1\rangle$), untrapped atoms ($|2\rangle$), excited state atoms ($|3\rangle$), and probe beam photons respectively, satisfying the usual bosonic commutation relations. 

The coupling coefficient, $g_{13}(\omega_k) = \frac{d_{13}}{\hbar}\sqrt{\frac{\hbar\omega_k}{2\epsilon_0}}$, where $d_{13}$ is the dipole moment for the $|1\rangle \rightarrow |3\rangle$ transition, is assumed to be approximately flat in the range of interest of our system.
If we assume that the detuning of the control beam from the excited state ($\Delta$) is large compared to all other frequencies in the system, including the kinetic energy due to the photon recoil, we can adiabatically eliminate the excited state, and obtain the following equations of motion for the Heisenberg operators,   
\begin{eqnarray}
i\dot{\hat{\psi}}_2 &=& H_2\hat{\psi}_2(x) - \Omega_C\tilde{E}(x) \label{psi_eom} \\
i\dot{\tilde{E}}(x) &=& H_E\tilde{E}(x)- \Omega_C^{*}\hat{\psi}_2(x)\label{E_eom} 
\end{eqnarray}
with $\tilde{E} = \hat{E}(x)e^{i(\omega_0 -\Delta)t)}$, $\Omega_C = \frac{g_{13}\Omega_{23}^*}{\Delta}e^{-ik_0x}\phi_1(x)$ and $H_2 = -\frac{\hbar}{2m}\nabla^2 -\frac{|\Omega_{23}|^2}{\Delta}$, and $H_E = -ic\frac{\partial}{\partial x} -(\omega_0 - \Delta) -\frac{|g_{13}|^2}{\Delta}|\phi_1(x)|^2$ with $\phi_1(x,t) \equiv \langle\hat{\psi}_1(x)\rangle$ representing the semiclassical wavefunction for the condensate atoms. In our calculation we have made the approximation that the condensate remains in a coherent state i.e. $\hat{\psi}_1(x) \approx \langle \hat{\psi}_1(x)\rangle \equiv \phi_1(x)$ . This is valid if the outcoupling is weak in the sense that the number of atoms outcoupled is small compared to the total number in the condensate. As strong atom-atom interactions would complicate the evolution of the quantum state of the condensate, we have assumed that it is dilute enough that these interactions can be ignored. The evolution of the condensate mode is then given by 
\begin{eqnarray}
i\dot{\phi}_1(x) &= (\frac{H_1}{\hbar} - \frac{g_{13}^2}{\Delta}\langle\tilde{E}^{\dagger}(x)\tilde{E}(x)\rangle)\phi_1(x) \nonumber \\ 
&-\Omega_C \langle\hat{E}^{\dagger}(x)\hat{\psi}_2(x)\rangle .
\end{eqnarray}

\begin{figure}
\includegraphics[width=8cm]{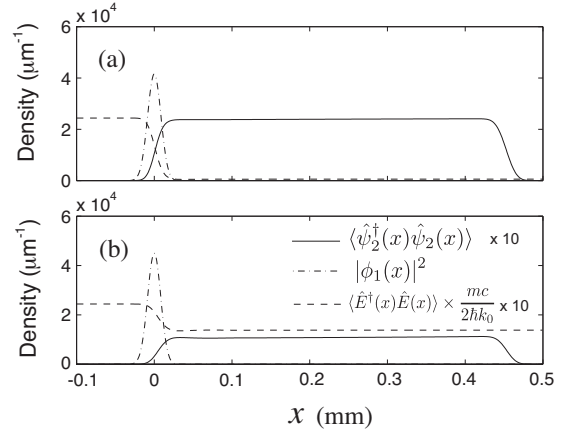}
\caption{\label{fig:density} Density of the condensate atoms $|\phi_0(x)|^2$, (dot dashed line), atom laser beam $\langle \hat{\psi}^{\dagger}_2(x)\hat{\psi}_2(x)\rangle$, (solid line), and the density of the probe optical field $\langle \hat{E}^{\dagger}(x)\hat{E}(x)\rangle$ multiplied by the ratio of the speed of light to the mean atomic speed, $\frac{m c}{2\hbar k_0}$ (dashed line), at $t= 40$ ms for (a) $\Omega_{23} = 1.5 \times 10^{8}$ rad s$^{-1}$ and (b) $\Omega_{23} = 0.75 \times 10^{8}$ rad s$^{-1}$ after $40$ ms. The probe beam and the atom laser beam have been multiplied by $100$ in order to fit them on the same scale as the condensate.}
\end{figure}

To analyse the dynamics of the system we expand the field operators in a mode basis, and solve for the dynamics of each of the mode functions, as described in \cite{squeezy2}. The equations of motion were integrated using a 4th order Runge Kutta algorithm with a cross propagation step for the optical field, using the numerical package XMDS \cite{xmds}. We have chosen realistic parameters for experiments with $^{87}$Rb atoms. Unless stated otherwise, we have set $m= 1.4 \times 10^{-25}$ kg, $g_{13} = 2.9 \times 10^{5}$ rad s$^{-1}$ m$^{\frac{1}{2}}$ and $\Delta = 10^{11}$ rad s$^{-1}$. We started with a condensate of $N=10^6$ atoms, initially in the ground state of a harmonic trap, with a trapping frequency of $5$ rad s$^{-1}$. The initial multimode quantum state of the probe optical field was chosen, in a plane wave basis, such that one mode (wave vector ${\bf k_p}$) was an arbitrary quantum state $|\gamma\rangle$, with a mean flux of $2.9\times 10^6$ photons/s,  with all other modes in the vacuum state. This mode was chosen such that the detuning from two-photon resonance was appropriate for an optimal Raman transition, and we assumed a geometry for the two optical fields such that the maximum possible momentum kick was transferred to the atoms, i.e. $|{\bf k_0} - {\bf k_p}| = 2k_0$. The initial quantum state of the untrapped atomic field was chosen as vacuum. Fig. \ref{fig:density} shows the density of the condensate, output beam, and probe field after $40$ ms of outcoupling for two different Rabi frequencies. 

In order to see how the quantum statistics are transferred in each case, we first need to define appropriate mode-matched amplitude and phase quadratures \cite{Armenia}. For the outcoupled atoms these are
\begin{eqnarray}
\hat{X}^{+} = \int_{x_1}^{x_2} L_{\psi}^*(x,t)\hat{\psi}(x,t) +  L_{\psi}(x,t)\hat{\psi}^{\dagger}(x,t) dx ,\\
\hat{X}^{-} = i\int_{x_1}^{x_2} L_{\psi}^*(x,t)\hat{\psi}(x,t) -  L_{\psi}(x,t)\hat{\psi}^{\dagger}(x,t) dx ,
\end{eqnarray}
while for the transmitted light we define
\begin{eqnarray}
\hat{Y}^{+} = \int_{x_1^{\prime}}^{x_2^{\prime}} L_{E}^*(x,t)\hat{E}(x,t) +  L_{E}(x,t)\hat{E}^{\dagger}(x,t) dx , \\
\hat{Y}^{-} = i\int_{x_1^{\prime}}^{x_2^{\prime}} L_{E}^*(x,t)\hat{E}(x,t) -  L_{E}(x,t)\hat{E}^{\dagger}(x,t) dx ,
\end{eqnarray}
where $L_{\psi}(x,t)$ and $L_{E}(x,t)$ are arbitrary modes which we choose as plane waves with appropriate wave length to best match the modes of the outcoupled atoms and transmitted light respectively, and are normalised on the interval of integration. We choose $x_1$ and $x_2$ to be points in the path of the atom laser beam, and $x_1^{\prime}$ and $x_2^{\prime}$ to be downstream of the condensate such that the light operators and atomic operators can be correlated at the same time, i.e. $x_1^{\prime} = \frac{c}{v_{atom}}x_1$, $x_2^{\prime} = \frac{c}{v_{atom}}x_2$, where $v_{atom}= \frac{\hbar 2 k_0}{m}$ is the mean speed of the atoms. In practice this would be impractical as $\frac{c}{v_{atom}}\approx 10^{11}$, and it would be more practical to detect the quadrature of the light and store it for later comparison with the atomic quadratures. We use the equal time definition here for convenience. The commutation relations give $V(\hat{X}^{+})V(\hat{X}^{-}) \geq 1$, $V(\hat{Y}^{+})V(\hat{Y}^{-}) \geq 1$.

\begin{figure}
\includegraphics[width=7cm]{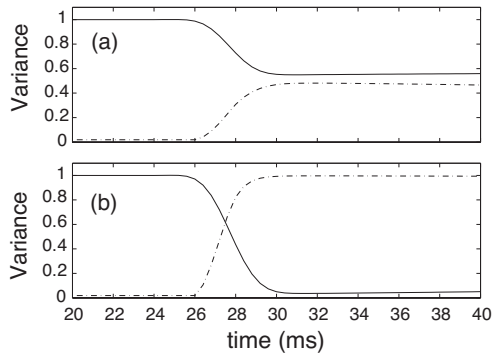}
\caption{\label{fig:variances} Variances of the amplitude quadratures for the atom laser beam (solid line), and probe beam (dot-dashed line) for (a) $\Omega_{23} = 0.75 \times 10^{8}$ rad s$^{-1}$, and (b) $\Omega_{23} = 1.5 \times 10^{8}$ rad s$^{-1}$. The initial state of the probe beam was an amplitude squeezed state with a squeezing parameter $r =2.0$. In case (b), the squeezing is almost completely transferred to the atom laser beam, which destroys the squeezing in the optical beam. In case (a), the squeezing is only partially transferred, and some squeezing remains in the optical beam. }
\end{figure}

Figure (\ref{fig:variances}) shows the variance of the amplitude quadratures for the atom laser beam ($\hat{X}^+$) and probe beam ($\hat{Y}^+$) versus time for the two cases shown in Figure \ref{fig:density}. The initial state of the optical field is chosen to be an amplitude squeezed state,  with $V(\hat{Y}^{\pm}(t=0)) = e^{\mp 2r}$ with the squeezing parameter $r = 2.0$.  In case (a), the squeezing is almost completely transferred to the atom laser beam, which destroys the squeezing in the optical beam. In case (b), the squeezing is only partially transferred, and some squeezing remains in the optical beam. The latter case is reminiscent of a 50/50 beam splitter.

\begin{figure}
\includegraphics[width=8cm]{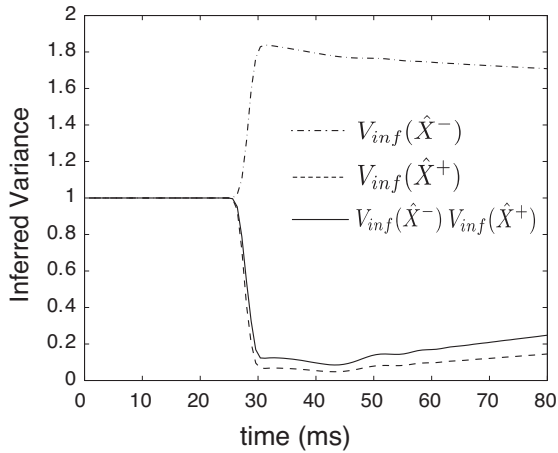}
\caption{\label{fig:Vinf} $V_{\mbox{inf}}(\hat{X}^-)$ (dot-dashed line), $V_{\mbox{inf}}(\hat{X}^+)$, (dashed line), and $V_{\mbox{inf}}(\hat{X}^+)V_{\mbox{inf}}(\hat{X}^-)$ (solid line) for $\Omega_{23} = 0.75 \times 10^{8}$ rad s$^{-1}$, and an initial amplitude squeezed optical state with the squeezing parameter $r = 2.0$.}
\end{figure}

To detect the entanglement between the probe beam and the atom laser beam we use the Einstein-Podolsky-Rosen (EPR) criterion of Reid and Drummond \cite{EPR}, who defined inferred quadratures which allowed for measurements that would seemingly violate the Heisenberg uncertainty principle, which is at the heart of the EPR paradox \cite{Albert}. In terms of measurable quantities, we consider the inferred variances $V_{\mbox{inf}}(\hat{X}^{\pm}) = V(\hat{X}^{\pm}) - \frac{[V(\hat{X}^{\pm}, \hat{Y}^{\pm})]^2}{V(\hat{Y}^{\pm})}$, where $V(\hat{X}^{\pm}, \hat{Y}^{\pm}) = \langle \hat{X}^{\pm}\hat{Y}^{\pm}\rangle - \langle\hat{X}^{\pm}\rangle\langle\hat{Y}^{\pm}\rangle$, resulting from the optimisation of a linear inference procedure. Quantitatively, $V_{\mbox{inf}}(\hat{X}^{+})V_{\mbox{inf}}(\hat{X}^{-})<1$ (similarly for the $Y^{\pm}$ quadratures) is then the requirement for entanglement. Figure \ref{fig:Vinf} shows the product of the inferred variances versus time for the case when $\Omega_{23} = 0.75\times 10^{8}$ rad s$^{-1}$ and $r = 2.0$. $V_{\mbox{inf}}(\hat{X}^+)V_{\mbox{inf}}(\hat{X}^-)$ dips well below $1$, demonstrating that there is entanglement between the transmitted probe beam and the atom laser beam. The entanglement slowly decreases (as did the squeezing in figure \ref{fig:variances}) due to two effects. The first effect is the depletion of the condensate, which changes the effective Rabi frequency between the optical field and the atomic field. This could be fixed by slowly increasing the power in the control beam ($\Omega_{23}$), or reducing the detuning. The second effect is due to the phases of the transmitted probe beam and atom laser beam drifting relative to their respective local oscillators, such that the quadratures being measured are no longer exactly orthogonal. This effect is due to the energy shift arising from the coupling process, and decreases as the intensity of the probe beam is decreased. 
We can compare our results to the case of an optical beam splitter with an amplitude squeezed beam input on one port, and vacuum input at the other port. In this case, with $r=2.0$, $V_{\mbox{inf}}(\hat{X}^{+}) = \frac{2e^{-2r}}{1+e^{-2r}} \approx 0.036$, $V_{\mbox{inf}}(\hat{X}^{-}) = \frac{2e^{2r}}{1+e^{2r}} \approx 1.96$, and $V_{\mbox{inf}}(\hat{X}^{+})V_{\mbox{inf}}(\hat{X}^{-}) = \frac{4}{2 + e^{2r}+e^{r}+e^{-2r}} \approx 0.071$. These compare quite well to the values which correspond to maximum entanglement in our system $V_{\mbox{inf}}(\hat{X}^+) \approx 0.048$, $V_{\mbox{inf}}(\hat{X}^-) \approx 1.83$, and $V_{\mbox{inf}}(\hat{X}^+)V_{\mbox{inf}}(\hat{X}^-) \approx 0.085$, indicating that our system behaves almost as an ideal beam splitter. 

{\it Experimental Considerations:}  For optimal squeezing in the atomic beam, the quantum efficiency of the outcoupling (ie. number of outcoupled atoms per probe beam photon passing through the condensate) must approach one, and for optimal entanglement between the probe beam and the atom laser beam we require the quantum efficiency to be one half. The parameter that governs the efficiency of the quantum state transfer is the effective Rabi frequency $\Omega_{\mbox{eff}} =  \frac{g_{13}\Omega_{23}^*}{\Delta}\int \phi_1(x)dx$. Optimum quantum state transfer occurs when the time taken to leave the condensate is approximately equal to a quarter period Rabi oscillation, i.e. $\Omega_{\mbox{eff}} = \sqrt{\frac{m\omega_{trap}}{\hbar}}\frac{\hbar|{\bf k_0} - {\bf k_p}|\pi}{2m}$. For the parameters used in our simulations this gives $\Omega_{23} \approx 1.6 \times 10^{8}$ rad s$^{-1}$. We choose a large detuning in order to reduce the effects of spontaneous emission. For $^{87}$Rb at this detuning, this Rabi frequency translates to an intensity of approximately $50$ mW cm$^{-2}$. To obtain the appropriate atom-light coupling coefficient $g_{13}$, we assumed that the probe beam was focussed to a waist of $100$ $\mu$m.

The most stringent requirement on the experiment will be on the intensity of the probe beam. This will have to be very weak in order to maintain high quantum efficiency in the outcoupling without saturating the condensate. To meet this requirement, the total number of photons in the experiment will have to be less than the number of atoms in the condensate. The photon flux from a squeezed vacuum is $F_{photon} \approx B\sinh^2r$, where $B$ is the bandwidth of the transition we are interested in. The bandwidth of a Raman transition when used to outcouple an atom laser however is just the inverse of the drain time of the condensate, $B \approx 1/\tau_{drain}$ \cite{linewidth}. The condition that the total number of photons be less than the total number of atoms in the condensate gives us $N_{atoms} >> N_{photons} = F_{photon}\tau_{drain} = \sinh^2r$, which is easily satisfied for experimentally achievable squeezing. Because the bandwidth of the atom laser transition is so narrow ($\sim$ kHz) however, in order to transfer squeezing to the outcoupled atoms, squeezing at low frequencies will be needed. Optical squeezing below $500$ Hz has recently been achieved by McKenzie {\it et al.} \cite{kng}.
High frequency squeezing can be thought of as entanglement between photons at frequencies on either side of a carrier beam. It may be possible to use high frequency squeezing to obtain atom light entanglement in an atom laser by making the sidebands on one side of the carrier resonant with the atom laser transition, and observing entanglement between the atom laser and the photons at the other frequency.

This research was supported by the Australian Research Council. We are grateful to A. M. Lance, K. McKenzie, N. P. Robins and M. T. Johnson for useful discussions relevant to this research.


\begin{thebibliography}{99}
\bibitem{linewidth} In the weak outcoupling limit, the frequency spread of the atom laser, and hence the absorbed light, is Fourier limited to the drain time of the condensate such that $\Delta \omega\Delta \tau \approx 1$. This will be presented in a future paper.
\end{thebibliography}
\end{document}